\documentclass[twocolumn,prb,showpacs]{revtex4}
\usepackage{graphicx}%
\usepackage{dcolumn}
\usepackage{amsmath}
\usepackage{latexsym}
\draft

\begin{document}


\title{Direct test of defect mediated laser induced melting theory for
      two dimensional solids} 
\author{Debasish Chaudhuri\footnote{debc@bose.res.in} and Surajit Sengupta\footnote{surajit@bose.res.in}
}
\affiliation{
Satyendra Nath Bose National Centre for Basic Sciences, Block-JD, Sector-III, Salt Lake,Calcutta - 700098.
}
\date{\today}

\begin{abstract}
We investigate by direct numerical solution of appropriate renormalization flow 
equations, the validity of a recent dislocation unbinding theory for laser 
induced freezing/melting in two dimensions. The bare elastic moduli and 
dislocation fugacities which are inputs to the flow equations are obtained 
for three different 2-d systems (hard disk, inverse $12^{th}$ power and 
the Derjaguin-Landau-Verwey-Overbeek potentials) from a restricted Monte Carlo
simulation sampling only configurations {\em without} dislocations. We 
conclude that (a) the flow equations need to be correct at least up to 
third order in defect fugacity to reproduce meaningful results, (b) 
there is excellent quantitative agreement between our results and 
earlier conventional Monte Carlo simulations for the hard disk system and 
(c) while the qualitative form of the phase diagram is reproduced for 
systems with soft potentials there is some quantitative discrepancy which
we explain.  
\end{abstract}
\maketitle

\section{Introduction}
\label{intro}

\noindent
Examples of phase transitions mediated by the unbinding of defect pairs abound
in two dimensions. The quasi- longranged- order to disorder transition in the 
XY and planar rotor models\cite{kt,surajitxy}, 
the melting transition of a two dimensional solid\cite{sura-hdmelt},
the superconductor to normal phase transition in two dimensional 
Josephson junction arrays\cite{joseph}, the 
commensurate- incommensurate transition of the striped  phase of 
smectic liquid crystals on anisotropic substrates\cite{stripe}, and the more 
recent discovery of a defect mediated re-entrant freezing transition in 
two dimensional colloids in an external periodic potential \cite{chowdhury,wei}
are all understood within such defect unbinding theories. While the very first  
defect mediated  transition theory for the phase 
transition in the  XY-model by Kosterlitz and Thouless (KT)\cite{kt} enjoyed 
almost immediate acceptance and was verified in simulations\cite{surajitxy}
as well as experiments\cite{xy-expt-1,xy-expt-2}, defect mediated theories of 
two dimensional melting took a long time to gain general acceptance in the 
community\cite{fight}. There were several valid reasons for this reticence 
however. 

\vskip .1cm
\noindent
Firstly, as was recognized even in the earliest papers\cite{kthny1,kthny2} 
on this subject, 
the dislocation unbinding transition, which represents an instability 
of the solid phase, may always be pre-empted by a 
first order\cite{strand,nelson-dg} transition 
from a metastable solid to a stable liquid. Whether such a first order 
melting transition actually occurs or not depends on the temperature 
of instability $T_{KT}$; so that if the transition temperature 
$T_c < T_{KT}$ the unbinding of dislocations does not occur. Clearly,
neither this condition nor its converse can hold for all 2d systems
in general since 
$T_{KT}$ is a non-universal number which depends on the ``distance''
in coupling parameter space between the bare and the fixed point 
Hamiltonian and hence on the details of the interaction. 
Secondly, the renormalization group flow equations derived in all defect 
mediated theories to date are perturbative expansions in the defect density 
(fugacity) in the ordered phase. How fast does this perturbation series 
converge? Again, the answer lies in the position of the bare Hamiltonian 
in the coupling parameter space. For the planar rotor model\cite{kt,surajitxy},
past calculations show that next to leading order terms in the flow equations 
are essential to reproduce the value of the transition
temperature obtained in simulations\cite{surajitxy}.   
Thirdly, defect mediated transitions predict an essential singularity\cite{kt} 
of the correlation length at the transition temperature. This means that 
effects of finite size\cite{clock} would be substantial and may thoroughly 
mask the true thermodynamic
result. A rapid increase of the correlation length also implies that the 
relaxation time diverges as the transition temperature is approached -- 
critical slowing down. For the two dimensional solid, this last effect is 
particularly crucial since, even far from the transition, the motion of 
defects is mainly thermally assisted and diffusional and therefore slow. 
The equilibration of defect configurations\cite{dyn-2dmelt}
 is therefore often an issue 
even in solids of macroscopic dimensions. 

\vskip .1cm
\noindent
On the other hand, over the last few years it has been possible to test 
quantitatively some of the non-universal predictions of defect mediated 
theories of phase transitions using simulations of restricted 
systems\cite{surajitxy,sura-hdmelt,mylif}. A simulation of a system without 
defects is used to obtain the values
for the bare coupling constants which are then taken as inputs to the 
renormalization group equations for the appropriate defect unbinding theory 
to obtain quantities like the transition temperature. Needless to say, the 
simulated system does not undergo a phase transition and therefore problems
typically related to diverging correlation lengths and times do not occur.
Numerical agreement of the result of this calculation with that of 
unrestricted simulations or experiments is proof of the validity of the RG 
flow equations\cite{kt,kthny1,kthny2,sura-hdmelt}. 
This idea has been repeatedly applied in the past to analyze 
defect mediated phase transitions in the planar rotor model\cite{surajitxy}, 
two dimensional melting of hard disks\cite{sura-hdmelt} and the re-entrant 
freezing of hard disks in an external periodic potential \cite{mylif,myerrlif}.
The last system is particularly interesting in view of its 
close relation with experiments on laser induced re-entrant freezing transition 
in charge stabilized colloids \cite{chowdhury,wei} and this constitutes the 
subject of the present paper as well. 

\vskip .1cm
\noindent
In this paper we show in detail how restricted simulations of systems 
of particles interacting among themselves via a variety of interactions and
with a commensurate external periodic potential can be used to obtain phase 
diagrams showing the re-entrant freezing transition. The results obtained are 
compared to earlier unrestricted simulations for the same systems. Briefly
our results are as follows. Firstly, we observe that, as in an earlier 
study of the planar rotor model\cite{surajitxy}, next to leading order 
corrections to the
renormalization flow equations are {\em essential} to reproduce even the 
gross features of the phase diagram. Specifically, the re-entrant portion of 
the phase diagram can be reproduced {\em only} if such correction terms 
are taken into account. Secondly while we find almost complete agreement with 
earlier results for the hard disk system which has been studied most 
extensively, our phase diagram for the other forms of interaction is shifted 
with respect to the results available in the literature. This may mean either
of two things --- inadequacy of the RG theory used by us or finite size 
effects in the earlier results. 
Lastly, as a by product of our 
calculations, we have obtained the core energy for defects (dislocations) in 
these systems and studied its dependence on thermodynamic and potential 
parameters.          

\vskip .1cm
\noindent
The problem of re-entrant freezing transition of a system of interacting 
colloidal particles in a periodic potential has an interesting history 
involving experiments\cite{chowdhury,wei}, 
simulations\cite{jcdlvo,cdas1,cdas2,cdas3,lif-hd,lif-dlvo,lif-sd1,lif-sd2} 
and theory\cite{jay,frey}. 
In last couple of decades soft systems
like colloids have been studied extensively\cite{colbook} both for their
own sake and as typical toy models to study various important
condensed matter questions like structural and phase transitions through
experiments that allow real space imaging. Charged colloids
confined within two glass plates form a model 2-d system as 
the electrostatic force from the plates almost completely suppresses the 
fluctuations of colloids perpendicular to the plates, practically confining 
them to a 2-d plane. In their pioneering 
experiment Chowdhury\cite{chowdhury} {\em et. al.} imposed a simple static 
background potential which is periodic in one direction and constant in the 
other (except for an overall Gaussian profile of intensity- variation) by 
interfering two laser beams. This potential immediately induces a density 
modulation in the colloidal system. The potential minima are spaced to 
overlap with the close packed lines of the ideal lattice of the colloidal 
system at a given density. With increase in potential strength 
such a colloidal liquid has been observed to solidify. This is known as laser 
induced freezing (LIF). In a recent experiment\cite{wei} it has been shown 
that with further increase in potential strength, surprisingly, the solid 
phase re-melts to a modulated liquid. This phenomenon is known as re-entrant 
laser induced freezing (RLIF).
Qualitatively, starting from a liquid phase, the external periodic potential
immediately induces a density modulation, reducing fluctuations which 
eventually leads to solidification. Further increase in the amplitude of 
the potential reduces the system to a collection of decoupled 1-d strips. 
The dimensional reduction now {\em 
increases} fluctuations remelting the system.

\vskip .1cm
\noindent
The early mean field theories, namely, Landau theory\cite{chowdhury} 
and density functional theory\cite{jay} predicted a change from a first 
order to continuous transition with increase in
potential strength and failed to describe the re-entrant behavior, 
a conclusion seemingly confirmed by early experiments\cite{chowdhury} and 
some early simulations\cite{jcdlvo}.
Overall, the results from early simulations remained inconclusive
however, while one of them\cite{jcdlvo}
claimed to have found a tri-critical point at intermediate laser intensities 
and re-entrance, later studies refuted these results
\cite{cdas1,cdas2,cdas3}. All of these studies used the change in order
parameter and the maximum in the specific heat to identify the phase 
transition points. While the later studies\cite{cdas1,cdas2,cdas3} found 
RLIF for hard disks they reported 
laser induced freezing and absence of any re-entrant melting for the DLVO 
potential\cite{cdas3} in direct contradiction to experiments\cite{wei}.

\vskip .1cm
\noindent
Following the defect mediated disordering approach of Kosterlitz and Thouless
\cite{kt}(KT), Frey, Nelson and Radzihovsky\cite{frey}(FNR) proposed a detailed theory for the re-entrant transition based on the unbinding of dislocations
with Burger's vector parallel to the line of potential minima. This theory 
predicted RLIF and no tricritical point. The results of this work were in 
qualitative agreement with experiments\cite{wei} and provided a framework
for understanding RLIF in general. More accurate simulation studies on 
systems of hard disks\cite{lif-hd}, soft disks\cite{lif-sd1,lif-sd2}, 
DLVO\cite{lif-dlvo} etc. confirmed the re-entrant freezing-melting transition 
in agreement with experiments\cite{wei} and FNR theory\cite{frey}. In these 
studies the phase transition point was found from the crossing of 
Binder-cumulants\cite{bincu,lan-bin} of order parameters corresponding
to translational and bond- orientational order, calculated for various 
sub- system sizes. A systematic finite size scaling analysis\cite{lif-hd} of 
simulation results for the 2-d hard disk system in a 1-d modulating 
potential showed, in fact, several universal features consistent with the 
predictions of FNR theory. It was shown in these studies that 
the resultant phase diagram remains system size dependent and the 
cross- over to the zero field KTHNY melting\cite{kthny1,kthny2} 
plays a crucial role in understanding the results for small values of the 
external potential. 
While the data collapse and critical exponents were consistent with KT
theory for stronger potentials, for weaker potentials they match better 
with critical scaling\cite{lif-hd}. A common problem with all the simulation
studies might be equilibration with respect to dislocation movements along
climb (or even glide) directions. Also, non universal predictions, namely 
the phase diagram are difficult to compare because 
the FNR approach (like KT theory) is expressed in terms of the appropriate 
elastic moduli which are notoriously time-consuming to compute near a 
continuous phase transition. Diverging correlation lengths and times near
the phase transition point further complicate an accurate evaluation of the 
non universal predictions of the theory. 
 
\begin{figure}[t]
\begin{center}
\includegraphics[width=4.0cm]{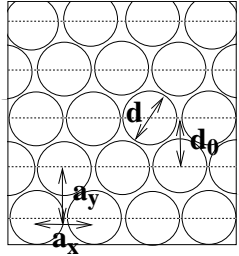}
\end{center}
\caption{This cartoon shows a typical 2-d system under consideration. 
$d$ is the length scale over which repulsive two body potentials are
operative. The dashed lines indicate minima of external modulating potential 
$\beta V(y)= -\beta V_0 \cos(2\pi y/d_0)$. $a_x=a_0$ is the lattice parameter 
fixed by the density $\rho$ and $a_y$ indicate the average separation 
between two layers along $y$-direction perpendicular to a set of close-packed 
planes. For a perfect triangular lattice $a_y=\sqrt{3}a_0/2$.
The modulating potential is commensurate with the lattice such that
$d_0=a_y$.}
\label{cartoon}
\end{figure}

\vskip .2cm
\noindent
We calculate the phase diagrams of three different 2-d systems 
with a 1-d modulating potential (see Fig.~\ref{cartoon}) following the
technique of restricted Monte Carlo simulations\cite{surajitxy, sura-hdmelt,
mylif}, to be discussed below. For the laser induced transition
we use this method to generate whole phase diagrams. 
We reject Monte Carlo moves which tend to distort an unit cell in a way which 
changes the local connectivity\cite{sura-hdmelt}. The percentage of moves thus 
rejected is a measure of the dislocation fugacity\cite{sura-hdmelt}. This, 
together with the elastic constants of the dislocation free lattice obtained 
separately, are our inputs (bare values) to the renormalization flow 
equations\cite{frey} to compute the melting points and hence the phase 
diagram. Our results (Fig.~\ref{hd-phdia},\ref{dlvo-phdia},\ref{soft-phdia}) 
clearly show a modulated liquid (ML) $\to$ locked floating solid (LFS) $\to$ 
ML re-entrant transition with increase in the amplitude ($V_0$) of the 
potential. In general, we find,  the predictions of FNR theory to be valid. 
\vskip .2cm

\noindent
In section \ref{method} we first briefly discuss the FNR theory and then 
go on to show in detail the restricted simulation scheme used by us to 
obtain the various quantities required to calculate the phase diagram. In  
section \ref{result} we give the simulation results. We describe, 
in detail, the various quantities leading to the phase diagram for one of the
systems, viz. the hard disks\cite{jaster,sura-hdmelt}. Then 
we present the phase diagrams for the other two systems we study. We compare
our results with earlier simulations. Lastly, in section \ref{conclusion} we 
summarize our main results and conclude.

\vskip .2cm

\vskip .2cm

\noindent
\section{method}
\label{method}
\noindent
A cartoon corresponding to the systems considered for our study is given in 
Fig.~\ref{cartoon}. The elastic free energy of the solid is given in terms of
the spatial derivatives of the displacement field 
$\vec u(\vec r) = \vec r - \vec r_0$ with $\vec r_0$ being the lattice vectors 
of the undistorted reference triangular lattice. For a solid in presence of a 
modulating potential $\beta V(y)$ (Fig.~\ref{cartoon}) the
displacement mode $u_y$ becomes massive, leaving massless $u_x$ modes. After
integrating out the $u_y$ modes the free energy of the LFS 
may be expressed in terms of gradients of $u_x$ and 
elastic moduli\cite{frey}, namely, the Young's
modulus $K(\beta V_0,\rho)$ and shear modulus $\mu(\beta V_0,\rho)$,
\begin{equation}
 {\cal H}_{el} = \int dx dy \left[ \frac{1}{2} K\left(\frac{\partial u_x}{\partial x}\right)^2+ \frac{1}{2} \mu\left(\frac{\partial u_x}{\partial y}\right)^2\right]
\label{hamiltonian} 
\end{equation}
\vskip .2cm
\noindent
Similar arguments\cite{frey} show that among the three sets of low 
energy dislocations available in the 2-d triangular lattice, only those 
(type I) with Burger's vector parallel to the line of potential minima survive 
at large $\beta V_0$. Dislocations with Burger's vector pointing along the 
other two possible close-packed directions (type II) in the 2-d triangular 
lattice have larger energies because the surrounding atoms are forced to ride 
the crests of the periodic potential\cite{frey}. Within this set of 
assumptions, the system therefore shares the same symmetries as 
the XY model. Indeed, 
a simple rescaling of $x\to\sqrt{\mu}x$ and $y\to\sqrt{K}y$ leads this free
energy to the free energy of the XY-model with spin-wave stiffness 
$K_{xy}=\sqrt{K\mu}a_0^2/4\pi^2$ and spin angle $\theta=2\pi u_x/a_0$:
$$
{\cal H}_{el} = \int dx dy ~ \left[\frac{1}{2}K_{xy}(\nabla\theta)^2\right]
$$ 
This immediately leads to the identification of a vortex in XY model 
($\oint d\theta = 2\pi$) with a dislocation of Burger's vector $ \vec b = 
\hat i a_0$ ($\oint du_x = a_0$, $\hat i =$ unit vector along $x$- 
direction) parallel to the potential minima {\em i.e.} the dislocation of 
type I. The corresponding theory for phase transitions can then be recast as a  
KT theory\cite{kt} and is described in the framework of a two parameter 
renormalization flow for the spin-wave stiffness $K_{xy}(l)$ and the fugacity 
of type I dislocations $y'(l)$,
where $l$ is a measure of length scale as $l=\ln(r/a_0)$, $r$ being the 
size of the system. The flow equations are expressed in
terms of $x'=(\pi K_{xy}-2)$ and $y'=4\pi~exp(-\beta E_c)$ where $E_c$ is the
core energy of type I dislocations which is obtained from the dislocation 
probability\cite{sura-hdmelt,morf}.
Keeping  upto next to leading order terms in $y'$ 
the renormalization group flow equations\cite{amit,surajitxy} 
are,
\begin{eqnarray}
\frac{dx'}{dl}&=& -y'^2 - y'^2x' \nonumber \\ 
\frac{dy'}{dl}&=& -x'y' + \frac{5}{4}y'^3.
\label{floweq}
\end{eqnarray}
Flows in $l$ generated by the above equations starting from initial or ``bare''
values of $x'$ and $y'$ fall in two categories. If, as  $l\to\infty$,  $y'$ 
diverges, the thermodynamic phase is disordered (i.e. ML), while on the other 
hand if $y'$ vanishes, it is an ordered phase (a LFS)\cite{frey}. 
The two kinds of flows are demarcated by the {\em separatrix} 
which marks the phase transition point. For the linearized equations, that
keeps upto only the leading order terms in $y'$, the 
separatrix is simply the straight line $y' = x'$, whereas for the full 
non-linear equations one needs to calculate this 
numerically\cite{amit,surajitxy,sura-hdmelt}.   

\begin{figure}[t]
\begin{center}
\includegraphics[width=8.6cm]{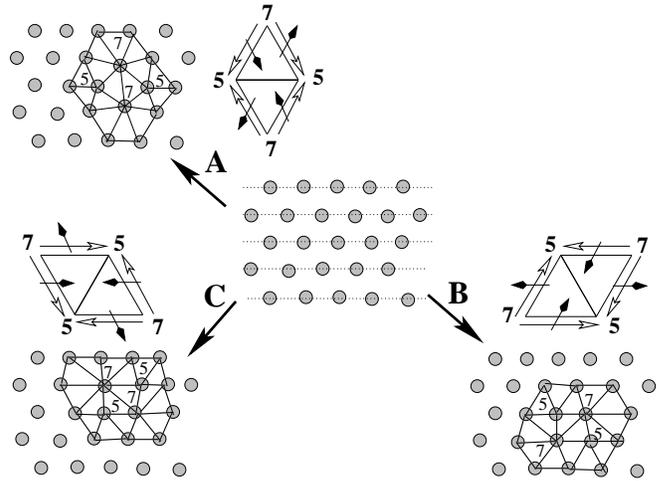}
\end{center}
\caption{
This diagram depicts all the possible dislocation generating moves that we 
reject. Starting from the triangular lattice shown in the centre (the dotted 
lines show the potential minima), in all, there can be three types of  
dislocation- pair generating moves shown as A, B \& C. The numbers $7$ 
and $5$ denote the positions of two types of disclinations having seven 
nearest neighbours and five nearest neighbours respectively. Only those bonds, 
which are necessary to show distortions due to the generation of disclination 
quartets, have been drawn. The rhombi near each of the distorted lattice 
denote the unit cells and open arrows from $7 \to 5$ show the direction of 
dislocation generating moves. The probabilities of these moves are $P_{m1}$(A),
$P_{m2}$(B) and $P_{m3}$(C). Corresponding Burger's vectors (filled arrows) 
are bisectors pointing towards a direction rotated counter-clockwise starting 
from $7 \to 5$ directions and are parallel to one of the lattice planes. 
Notice, the separation between Burger's vectors of a pair along the glide 
direction (parallel to the Burger's vectors) is a single lattice separation 
($a_0$) and within this construction it is impossible to draw Burger's loop 
that can generate non-zero Burger's vector. Depending on which of the two 
possible disclination pairs separate out any one dislocation- antidislocation 
pair will be formed.   
}
\label{dislo-cartoon}
\end{figure}

\vskip .2cm
\noindent
The bare numbers for $x'$ and $y'$ are 
relatively insensitive to system size since our Monte Carlo simulation does not
involve a  diverging correlation length associated with a phase transition. 
This is achieved as follows\cite{surajitxy,sura-hdmelt}. 
We monitor individual random moves of the particles in a system
and look for distortions of the neighboring unit cells. If in 
any of these unit cells the length of a next nearest neighbor bond 
becomes smaller than the nearest neighbor bond, the move is rejected. 
All such  moves generate disclination quartets and are shown in the  
Fig.~\ref{dislo-cartoon}. Notice that each of these moves break a nearest 
neighbour bond to build a new next nearest neighbour bond, in the process 
generating two $7$-$5$ disclination pairs. These are
the moves rejected in the restricted simulation scheme we follow. 
The probabilities of such bond breaking moves are however computed by 
keeping track of the number of such rejected moves. One has to keep 
track of all the three possible distortions of the unit rhombus  with 
measured probabilities $P_{mi}, i = 1,3$ (see Fig.~\ref{dislo-cartoon}). 
Each of these distortions involves four $7-5$ disclinations {\em i.e.} 
two possible dislocation- antidislocation pairs which, we assume, occur 
independently. For a free ($V_0 = 0$) two dimensional system dislocation 
core energy $E_c^t$ can be found through the relation\cite{morf}
\begin{equation}
\Pi = \exp(-\beta\,\, 2E_c^t)Z(\tilde K)
\label{ec0}
\end{equation}
where $\Pi=\sum_{i=1}^3 P_{mi}$ and $Z(\tilde K)$ is the ``internal partition function" incorporating all three types of degenerate orientations of 
dislocations, 
\begin{equation}
Z(\tilde K) = \frac{2\pi\sqrt{3}}{\tilde K/8\pi -1}
\left( \frac{r_{min}}{a_0}\right)^{2-\tilde K/4\pi}
I_0\left(\frac{\tilde K}{8\pi}\right)
\exp\left(\frac{\tilde K}{8\pi}\right)\nonumber
\end{equation}
where $I_0$ is a modified Bessel function, $\tilde K = \beta K a_0^2$ is a
dimensionless Young's modulus renormalized over phonon modes,
$a_0$ being the lattice parameter and $r_{min}$
is the separation between dislocation-antidislocation above which one counts 
the pairs. The above expression for $Z(\tilde K)$ and Eq.(\ref{ec0}) have been 
used previously in simulations\cite{sura-hdmelt,morf} of phase transitions of
2-d systems in absence of any external potential to find the dislocation core 
energy $E_c^t$.

\vskip .2cm

\noindent
The probabilities for occurrence of the dislocation pairs of a specific type
themselves $P_{di}$ (Fig.~\ref{dislo}) which are proportional to the square 
of the fugacities, can be computed easily. 
 The probability of dislocation pairs of type I is 
$P_{d1}=\frac{1}{2}(P_{m2}+P_{m3}-P_{m1})$ and that of type II is 
$P_{d2} = \frac{1}{4}(P_{m1}+P_{m2}+P_{m3} - 2 P_{d1}) = {P_{m1}}/{2}$.
The validity of the above expressions can be clearly seen from 
Fig.\ref{dislo-cartoon}.

\noindent
An argument following the lines of Fisher {\em et. al.}\cite{morf} 
shows that the dislocation probability (number density of dislocation pair 
per unit cell) for our system,
\begin{equation}
P_{d1} = \exp(-\beta\,\, 2 E_c) Z(\tilde K_{xy})
\label{fuga1}
\end{equation}
where $2E_c$ is the core energy and $Z(\tilde K_{xy})$ is the internal 
partition function of dislocation pair of type I (single orientation). 
\begin{eqnarray}
Z(\tilde K_{xy}) &=& \int_{r>r_{min}} \frac{d^2 r}{A_c} 
\exp\left[ -2\pi \tilde K_{xy} \log\left(\frac{r}{a_0}\right)\right] \nonumber \\ 
&=& \frac{2\pi}{\sqrt{3}}\,\, \frac{(r_{min}/a_0)^{2-2\pi \tilde K_{xy}}}{\pi \tilde K_{xy}-1}
\label{fuga2}
\end{eqnarray}
with $\tilde K_{xy} = \beta K_{xy}$ and $A_c = \sqrt3a_0^2/2$ being the area of an unit cell in the undistorted lattice. We choose $r_{min}=2a_0$. At this point
this choice is arbitrary. We give the detailed reasoning for this choice at the
end of section \ref{result}.
Eq.\ref{fuga1} and Eq.\ref{fuga2} straightaway yield the required core energy 
$E_c$. The corresponding fugacity contribution to RG flow equations 
(Eq.\ref{floweq}) is given via
\begin{equation}
y'=4\pi\sqrt{P_{d1}/Z(\tilde K_{xy})}
\label{y'}
\end{equation}

\noindent
In the above, the following assumption is, however, implicit. Once a 
nearest neighbor bond breaks and a potential dislocation pair is formed,
they separate with probability one\cite{kram-like}.
This assumption goes into the identity Eq.\ref{fuga1} as well as in 
Eq.\ref{ec0}\cite{sura-hdmelt}.
Taking the rejection ratios due to bond- breaking as the dislocation 
probabilities, as well, require this assumption\cite{lif-note-1}.

\vskip .2cm

\noindent
The same restricted Monte Carlo simulation can be used to find out the stress 
tensor, and the elastic moduli from the stress-strain curves.
The dimensionless stress tensor for a free ($V_0 = 0$) system is given 
by\cite{elast}, 
\begin{equation}
\beta \sigma_{\lambda\nu} {\rm d}^2 = -\frac{{\rm d}^2}{S}\left(-\sum_{<ij>}\left< \beta\frac{\partial \phi}{\partial r^{ij}}~\frac{r_\lambda^{ij}r_\nu^{ij}}{r^{ij}}\right> + N\delta_{\lambda\nu}\right)
\label{sts1}
\end{equation} 
where $i$, $j$ are particle indices and $\lambda$, $\nu$ denote directions $x$, $y$; $\phi(r^{ij})$ is the two- body interaction, $S/{\rm d}^2$ is the 
dimensionless area of the simulation box \cite{pot-impact}.

\noindent
\section{results and discussion}
\label{result}
\noindent
In this section we present the calculation of the phase diagram
for three different 2-d systems, namely hard disks, soft disks and 
a system of colloidal particles interacting via the DLVO 
(Derjaguin-Landau-Verwey-Overbeek)\cite{dlvo1,dlvo2} potentials.
We discuss, first, the calculation of the phase diagram for a two dimensional 
system of hard disks, in detail. 
The bulk system of hard disks where particles $i$ and $j$, in 2-d, 
interact via the potential $\phi(r^{ij}) = 0$ for $r^{ij} > {\rm d}$ 
and $\phi(r^{ij}) = \infty$ for $r^{ij} \leq {\rm d}$, where 
${\rm d}$ is the hard disk diameter and 
$r^{ij} = |{\bf r}^j - {\bf r}^i|$ the 
relative separation of the particles, is known 
to melt\cite{al,zo,web,jaster,sura-hdmelt} from a high 
density triangular lattice to an isotropic liquid with a narrow 
intervening hexatic phase\cite{kthny1,kthny2,jaster,sura-hdmelt}. 
The hard disk free energy is entirely entropic in 
origin and the only thermodynamically relevant variable is the number density   
$\rho = N/V$ or the packing fraction $\eta = (\pi/4) \rho {\rm d}^2$.
Simulations\cite{jaster}, experimental\cite{colbook} and 
theoretical\cite{rhyzov} studies of hard 
disks show that for $\eta > .715$ the system exists as a triangular 
lattice which transforms to a liquid below $\eta = .706$. The small 
intervening region contains a hexatic phase predicted by 
the KTHNY theory\cite{kthny1,kthny2} of 2-d melting. 
 Apart from being easily 
accessible to theoretical treatment\cite{hansen-macdonald}, experimental systems
with nearly ``hard'' interactions viz. sterically stabilized 
colloids\cite{colbook} are available.  
\begin{figure}[t]
\begin{center}
\includegraphics[width=8.6cm]{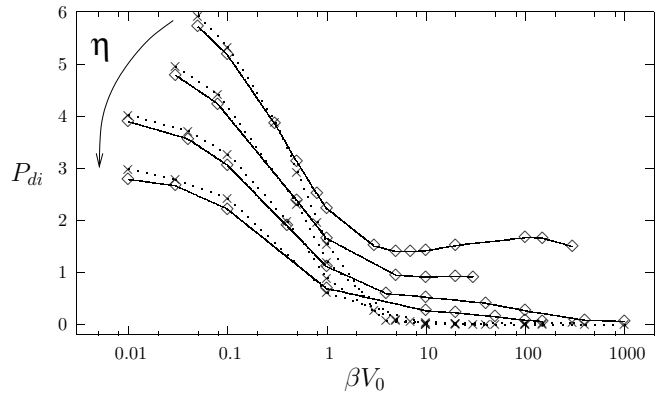}
\end{center}
\caption{
In this plot the {\Large $\diamond$} 
symbols correspond to $P_{d1}$, the probability for type I dislocations and 
the {\large $\times$} symbols to  $P_{d2}$ the probability for type II 
dislocations obtained from the $P_{mi}$ (see text and Fig.\ref{dislo-cartoon})
for various $\eta$ values, arrow denoting the direction of increasing
$\eta$($=.69,.696,.7029,.71$). $P_{di}$ for $i=1,2$ are expressed
in units of $10^4$.
These probabilities are plotted against the  potential strength 
$\beta V_0$. Note that for $\beta V_0 > 1$, the probability for type I
dislocations is larger than that of type II. The dots and solid lines are
only guides to eye.
}
\label{dislo}
\end{figure}

In presence of a periodic external potential, the only other energy scale 
present in the system is the relative potential\cite{dielec}
strength $\beta V_0$. 
If the modulating potential is commensurate with the spacing 
between close- packed lines, the elastic free energy of this system in it's
solid phase follows Eq.\ref{hamiltonian} and the corresponding renormalization 
flow equations are given by Eq.\ref{floweq}.

\vskip .2cm
\noindent
We obtain the bare $y'$ and $x'$ from Monte Carlo simulations of
$43\times 50=2150$ hard disks and use them as initial 
values for the numerical solution of Eqs.\,(\ref{floweq}). 
The Monte Carlo simulations for hard disks is done in the usual\cite{frenkel}
way viz. we perform individual
random moves of hard disks after checking for overlaps with neighbours.
When a move is about to be accepted, however, we  
look for the possibility of bond breaking as described in the previous section 
(Fig.\ref{dislo-cartoon}). We reject any such move and the rejection ratios
for specific types of bond breaking moves give us the
dislocation probabilities of type I and II, separately (Fig.\ref{dislo}).
From Fig.\ref{dislo} it is clear that the probability of type
II  dislocations {\em i.e.} $P_{d2}$ drops down to zero for all packing 
fractions at higher 
potential strengths $\beta V_0$. The external potential suppresses formation
of this kind of dislocations. For small $\beta V_0$ on the other hand, the 
probabilities of type I and type II dislocations are roughly the same.
This should be a cause of concern since we neglect the contribution of type
II dislocations for {\em all} $\beta V_0$. We comment on this issue later
in this section.

\vskip .1cm
\noindent
Using Eq.\ref{y'} and Eq.\ref{fuga2} along
with the identity $r_{min}=2 a_0$ gives us the initial value $y'_0$ to be
used in renormalization flow Eq.\ref{floweq} provided we know 
$\tilde K_{xy}$. Again $K_{xy}$ gives $x'$ straightaway. To obtain that we
need to calculate the Young modulus $K$ and shear modulus $\mu$.
 
\begin{figure}[t]
\begin{center}
\includegraphics[width=7.0cm]{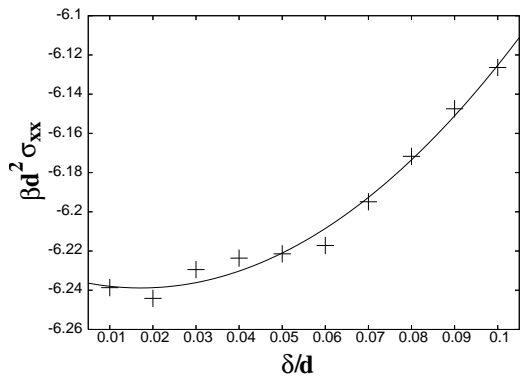}
\end{center}
\caption{Plot of $\beta d^2\sigma_{xx}$ vs. $\delta/{\rm d}$ at a strain value $\epsilon_{xx}=.02$ for packing fraction $\eta = .7029$ and potential strength $V_0=1$. A second order polynomial fit (solid line) gives $\lim_{\delta\to 0}\beta d^2\sigma_{xx}=-6.23$ .
}
\label{ydel}
\end{figure}
\begin{figure}[t]
\begin{center}
\includegraphics[width=7.0cm]{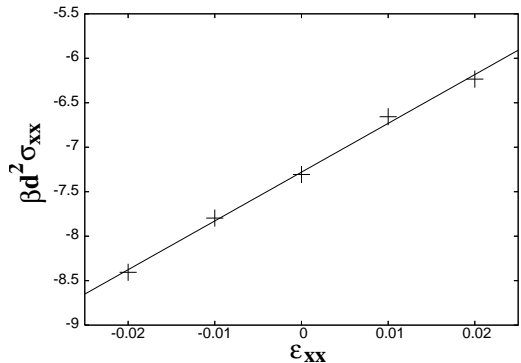}
\end{center}
\caption{A typical stress-strain curve used to obtain the Young modulus from a linear fit (solid line). The graph is plotted at $\eta = .7029,~ V_0=1.0$. The fitted Young's modulus $\beta K {\rm d}^2=54.84$.
}
\label{ymod}
\end{figure}
\begin{figure}[t]
\begin{center}
\includegraphics[width=7.0cm]{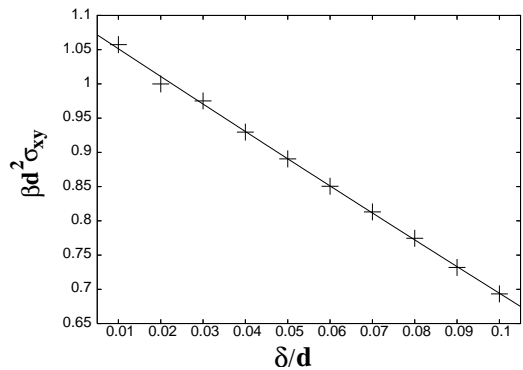}
\end{center}
\caption{Plot of $\beta d^2\sigma_{xy}$ vs. $\delta/{\rm d}$ at strain value $\epsilon_{xy}=.08$ at the packing fraction $\eta = .7029$ and potential strength $V_0=1$. A second order polynomial fit (solid line) gives $\lim_{\delta\to 0}\beta d^2\sigma_{xy}=1.092$ . 
}
\label{shdel}
\end{figure}
\begin{figure}[t]
\begin{center}
\includegraphics[width=7.0cm]{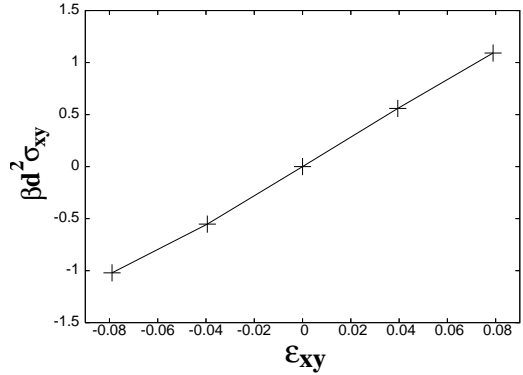}
\end{center}
\caption{
A typical stress-strain curve used to obtain shear modulus from a linear
fit (solid line). The graph is plotted at $\eta = .7029,~ V_0=1.0$. The fitted shear modulus $\beta \mu {\rm d}^2=13.53$.
}
\label{shmod}
\end{figure}
\noindent
 
Let us  again go
back to Eq.\ref{sts1}, the expression for stress tensor. For hard disk 
potentials the derivative $\partial\phi/\partial r^{ij}$ becomes a 
Dirac delta function and the expression for stress can be recast 
into\cite{elast}
\begin{equation}
\beta \sigma_{\lambda\nu} {\rm d}^2 = -\frac{{\rm d}^2}{S}\left(\sum_{<i,j>}\left< \frac{r_\lambda^{ij}r_\nu^{ij}}{r^{ij}}\delta(r^{ij}-d)\right> + N\delta_{\lambda\nu}\right)
\label{sts2}
\end{equation}
The presence of Dirac delta function $\delta (r^{ij}-d)$ in the above 
expression requires 
that the terms under the summation contribute, strictly, when two 
hard disks touch each other {\em i.e.} $r^{ij}\equiv r = \sigma$.
In practice, we implement this, by  adding the
terms under summation when each pair of hard disks come within a small 
separation $r = \sigma + \delta$. We then find 
$\beta \sigma_{\lambda\nu} {\rm d}^2$ as function 
of $\delta$ and fit the curve to a second order polynomial.
Extrapolating to the $\delta \to 0$ limit obtains the 
value of a given component of stress tensor at each strain value 
$\epsilon_{\lambda\nu}$\cite{elast}. 

\vskip .1cm
\noindent
For completeness, now we show how we calculate the two relevant 
stress-tensors : $\sigma_{xx}$ at a given longitudinal strain $\epsilon_{xx}$
in Fig.~\ref{ydel} and $\sigma_{xy}$ for a shear strain $\epsilon_{xy}$
in Fig.~\ref{shdel}. We thus calculate the stress at each value of strain 
and from the slopes of stress-strain curves find out the bare Young-modulus 
$\beta K {\rm d}^2$ (Fig.~\ref{ymod}) and shear-modulus 
$\beta \mu {\rm d}^2$(Fig.~\ref{shmod}). To obtain the relevant elastic moduli
we give first an elongational strain in $x$- direction which is parallel to the 
direction of potential minima to obtain $K$ and subsequently a shear 
in the same direction to obtain $\mu$. Any strain that forces the system to 
ride potential hills will give rise to massive displacement modes which do 
not contribute to elastic theory.

\begin{figure}[t]
\begin{center}
\includegraphics[width=7.0cm]{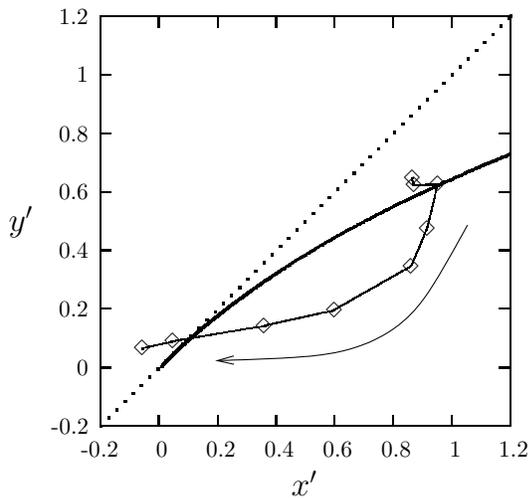}
\end{center}
\caption{
The initial values of $x'$ and $y'$ obtained from the
elastic moduli and dislocation probability
at $\eta=.7029$ plotted in $x'-y'$ plane. The line connecting the points
is a guide to eye. The arrow shows the direction of increase in
$\beta V_0$($=.01, .04, .1, .4, 1, 4, 10, 40, 100$).
The dotted line denotes the separatrix ($y'=x'$) incorporating only the
leading order term in KT flow equations. The solid curve is the separatrix
when next to leading order terms are included. In
$l\to\infty$ limit any initial value below the separatrix flows to $y'=0$ line
whereas that above the separatrix flows to $y'\to\infty$. The intersection
points of the line of initial values with
the separatrix gives the phase transition points. The plot shows a
freezing transition at $\beta V_0=.1$ followed by a melting at $\beta V_0=30$.
}
\label{flow}
\end{figure}

\begin{figure}[t]
\begin{center}
\includegraphics[width=7.0cm]{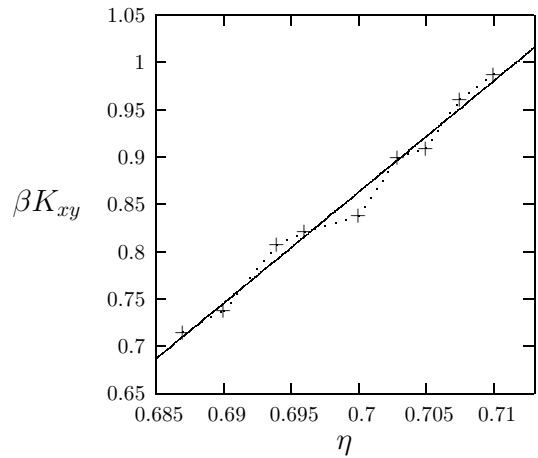}
\end{center}
\caption{For hard disk system $\beta K_{xy}$ varies linearly with
$\eta$. Data plotted at $V_0 = 1$. The solid line is a linear fit to the form $f(x)=a+b x$ with $a=-7.37$ and $b=11.76$. At each $V_0$ the error in 
$K_{xy}$  determines the error in $\eta$: $\delta \eta/\eta = 
|1 + a/\eta b|  (\delta K_{xy}/K_{xy})$.
} 
\label{error1}
\end{figure}
\begin{figure}[t]
\begin{center}
\includegraphics[width=8.6cm]{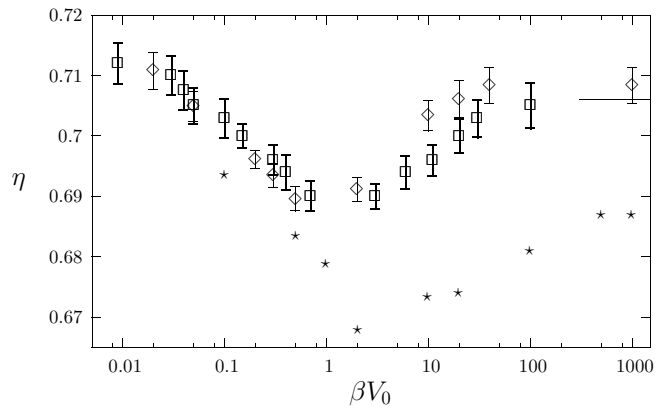}
\end{center}
\caption{The phase diagram of the hard disk system in the presence of a
1-d, commensurate, periodic potential in the packing fraction ($\eta$) -
potential strength ($\beta V_0$) plane. The points denoted by $\Box$ 
correspond to our RG calculation using the techniques described in this paper.
The points denoted by $\Diamond$\cite{lif-hd} and $\ast$\cite{cdas3} are 
taken from earlier simulations. The vertical bars denote estimate of error. 
Our data clearly matches with Ref[7].The horizontal line at $\eta = .706$ 
denotes the calculated asymptotic phase transition point at $\beta V_0=\infty$.} 

\label{hd-phdia}
\end{figure}

\vskip .1cm
\noindent
From these elastic moduli we get the `bare' $K_{xy}$ (and hence 
$x'_0 = \pi K_{xy}-2$, see section~\ref{method}). This is also 
required to complete the computation of $y'_0$.
In Fig.~\ref{flow} we have plotted $x'_0$ and $y'_0$ the bare values of 
$x'$ and $y'$ for various potential strengths $\beta V_0$ at
packing fraction $\eta=.7029$  
along with the separatrices for the linearized and the non-linear flow 
equations (Eq. \ref{floweq}). The line of initial conditions is seen to 
cross the non-linear separatrix twice (signifying re-entrant behaviour)  
while crossing  the  corresponding linearized separatrix only once at high 
potential strengths. 
The phase diagram (Fig.~\ref{hd-phdia}) is obtained by computing the 
points at which the line of 
initial conditions cut the non-linear separatrix using a simple 
interpolation scheme.  
It is interesting to note that within a linear theory the KT flow 
equations {\em fail to predict a RLIF transition}. 
Performing the same calculation for different packing fractions $\eta$
we find out the whole phase diagram of RLIF in the $\eta$- $\beta V_0$
plane. 
\vskip .1cm 

\noindent
The numerical errors in the phase diagram are calculated as follows.
The quantity $\beta K_{xy}$ varies linearly with $\eta$ at all potential
strengths. Therefore the numerical error in $\eta$ is proportional to
the error in $\beta K_{xy}$ (see Fig. \ref{error1}). Using all these we
obtain the RLIF phase diagram for hard disk systems (Fig.\ref{hd-phdia}).


\vskip .1cm
\noindent
Further, comparing with previous computations\cite{lif-hd, cdas3} of the 
phase diagram for this system (also shown in Fig.~\ref{hd-phdia})
we find that, within error- bars, our results agree at all values 
of $\eta$ and $\beta V_0$ with the results of W. Strepp {\em et. al.}
\cite{lif-hd}. Whereas, in numerical details, they disagree with the results 
of C. Das {\em et. al.}\cite{cdas3}, though even these results show RLIF and 
are in qualitative agreement with ours. 
This validates both our method and the quantitative predictions of 
Ref. \cite{frey}. The effect of higher order terms in 
determining non-universal quantities has been pointed out 
earlier\cite{surajitxy} for the planar rotor model but in the present case 
their inclusion appears to be crucial.   
Nevertheless, we expect our procedure to break down 
in the $\beta V_0 \to 0$ limit where effects due to the cross-over 
from a KT to a KTHNY\cite{kthny1,kthny2} transition at $\beta V_0 = 0$ become 
significant. Indeed, as is evident from Fig.~\ref{dislo} for $\beta V_0 < 1$ 
the dislocation probabilities of both type I and type II 
dislocations are similar\cite{lif-note-2} and the assumptions of FNR theory and
our process (which involves only type I 
dislocations) need not be valid at small potential strengths. 
This fact is also supported by  
Ref.\cite{lif-hd} where it was shown that though at $\beta V_0 = 1000$ the 
scaling of susceptibility and order parameter cumulants gave good data 
collapse with values of critical exponents close to FNR predictions, at 
$\beta V_0 = .5$, on the other hand, ordinary critical scaling gave better data collapse than
the KT scaling form, perhaps due to the above mentioned crossover effects. 
In the asymptotic limit of $\beta V_0\to\infty$ the system
freezes above $\eta = .706$ which was determined from a separate simulation 
in that limit. This number is very close to the earlier 
value $\eta\sim.71$ quoted in Ref.\cite{lif-hd}. As expected, the 
freezing density in the $\beta V_0 \to \infty$ limit is lower than 
the value without the periodic potential {\em i.e.} $\eta \simeq .715$.  
\vskip .2cm
\begin{figure}[t]
\begin{center}
\includegraphics[width=8.6cm]{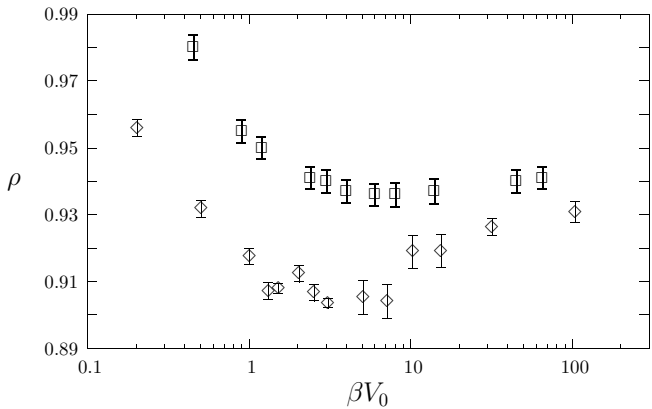}
\end{center}
\caption{ Phase diagram for soft disks:
$\Box$   denote our calculation,
$\Diamond$ indicate earlier simulation data\cite{lif-sd1,lif-sd2}.
The vertical lines are the error- bars. 
}
\label{soft-phdia}
\end{figure}
 \begin{figure}[t]
\begin{center}
\includegraphics[width=8.6cm]{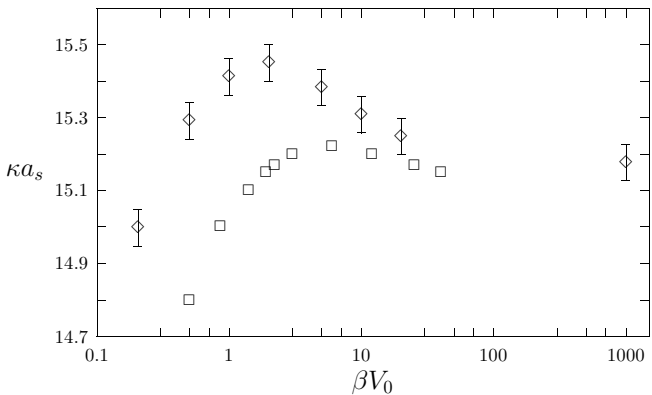}
\end{center}
\caption{ Phase diagram for particles interacting via the DLVO potential.
$\Box$ denote our calculation, $\Diamond$ show the earlier simulation 
data\cite{lif-dlvo}. The vertical lines are the error- bars. Error bars
in our calculation being smaller than the symbol size are not shown.
}
\label{dlvo-phdia}
\end{figure}
\vskip .1cm
\noindent
In a similar fashion it is possible to find out phase diagrams of
any 2-d system in presence of external modulating potential commensurate 
with the density of the system. We illustrate this by calculating similar
phase diagrams for two other systems, viz. soft disks and the DLVO system.
The soft disks interact via the potential :
$$
\phi(r) = \frac{1}{r^{12}} 
$$
where $r$ denotes the separation between particles. In simulations,
the cutoff distance is chosen to be $r_c=2$ above which the particles are
assumed to be non- interacting. Apart from the external
potential strength $\beta V_0$ the relevant thermodynamic quantity is the
number density $\rho = N/L_xL_y$. In finding `bare'
elastic moduli from restricted simulations the stress is calculated from
the Eq.\ref{sts1}. As this expression does not involve any Dirac delta 
functions unlike the case of hard disks, we do not require any fitting like
Figs.\ref{ydel},\ref{shdel} to obtain the stresses. The elastic moduli are
again found from stress- strain curves like Figs.\ref{ymod},\ref{shmod}. The
dislocation fugacity of type I is calculated from rejection ratio of 
dislocation generating moves. All these, at a given $\rho$ value
generate the initial conditions $x'_0$ and $y'_0$ in RG flow diagrams.
The crossing of these initial conditions with the separatrix found
from Eq.\ref{floweq} gives the phase transition points. The phase diagram 
is plotted and compared with phase diagram from earlier 
simulations\cite{lif-sd1,lif-sd2} in 
Fig.\ref{soft-phdia}. The error bar in $\rho$ is found from the
error in $K_{xy}$, as $K_{xy}$ varies linearly with $\rho$, through the 
relation $\delta \rho/\rho = |1 + a/\rho b|  (\delta K_{xy}/K_{xy})$.
The quantities $a$ and $b$ are found from linear fitting (of form $a+bx$) of
$K_{xy}$ vs. $\rho$ curve, at any given $\beta V_0$.
The phase diagram (Fig. \ref{soft-phdia}) again clearly shows 
re-entrance (RLIF). This is in qualitative agreement with earlier 
simulations\cite{lif-sd1,lif-sd2} (see Fig.\ref{soft-phdia}). 
\vskip .2cm
\noindent 
For charge stabilized colloids the inter-particle potential that operates is
approximately given by the DLVO potential \cite{dlvo1,dlvo2}:
$$
\phi(r) = \frac{(Z^\ast e)^2}{4\pi\epsilon_0\epsilon_r}\left( \frac{exp(.5\kappa {\rm d})}{1 + .5\kappa {\rm d}}\right)^2 \frac{exp(-\kappa r)}{r}
$$
where $r$ is the separation between two particles,
d is the diameter of the colloids, $\kappa$ is the inverse Debye 
screening length, $Z^\ast$ is the amount of effective surface charge and
$\epsilon_r$ is the dielectric constant of the water in which the colloids 
are floating. In order to remain close to experimental situations and to
be able to compare our phase diagram with the simulations of 
Strepp {et. al.}\cite{lif-hd} we use $T=293.15 K$, 
d$=1.07\mu m$, $ Z^\ast = 7800$, $\epsilon_r=78$. 
In experiments, the  dimensionless inverse Debye screening length
$\kappa a_s$ can be varied either by changing $\kappa$ through the 
change in counter-ion concentration or by changing $a_s$ by varying 
density\cite{bch-fr}.  We perform the
restricted Monte- Carlo simulation as described in section \ref{method}. In 
simulations we vary $\kappa$  and keep the particle spacing in
ideal lattice $a_s=2.52578 \mu m$ (density) fixed. 
Further, we use a cut- off radius $r_c$ such that, $\phi(r>r_c) = 0$ where
$r_c$ is found from the condition $\beta\phi(r_c)=.001$. We find out phase 
transition points (in $\kappa a_s$) at different external potential strengths 
$\beta V_0$ in the same fashion as described earlier.
The bare renormalizable quantities $x'_0$ and $y'_0$ are
found from restricted Monte- Carlo simulations for various $\beta V_0$ at 
each $\kappa a_s$.  The phase transition points are calculated from the 
intersection of these initial conditions with the separatrix found from 
Eq.\ref{floweq}. Thus we obtain 
the phase diagram in $\kappa a_s-\beta V_0$ plane (Fig. \ref{dlvo-phdia}).
$\beta K_{xy}$ varies linearly with $\kappa a_s$ and the error in 
$K_{xy}$ generates the error in $\kappa a_s$  (Fig.\ref{dlvo-phdia})
through the relation 
$\delta (\kappa a_s)/(\kappa a_s) = |1 + a/b \kappa a_s|(\delta K_{xy}/K_{xy})$.
The quantities $a$ and $b$ are found from linear fitting (of form $a+bx$) of
$K_{xy}$ vs. $\kappa a_s$ curve, at any given $\beta V_0$.
Though there is a quantitative mismatch between our data and that of Strepp
{\em et. al.}\cite{lif-dlvo}, our data shows a clear region in 
$\kappa a_s$ (between $15.1$ and $15.2$) where we obtain re-entrance (RLIF).
This is in contrast to the simulated phase diagram of C. Das 
{\em et. al.}\cite{cdas3}, where they observe absence of re-entrance at high
field strengths. We do not plot their data as the parameters these
authors used are not the same as the ones used in Fig.\ref{dlvo-phdia}.

\vskip .2cm

\noindent
It is interesting to note that, with increase in
range of two- body interaction potentials the depths of re-entrnace (in $\eta$,
$\rho$ or $\kappa a_s$) decreases. This is again in agreement with the 
understanding that, the re-entrant melting comes about due to decoupling of the
1-d trapped layers of particles that reduces the effective dimensionality 
thereby increasing fluctuations. With the increase in range of the interacting 
potentials this decoupling gets more and more suppressed, thereby reducing
the region of re-entrance.

\begin{figure}[t]
\begin{center}
\includegraphics[width=8.6cm]{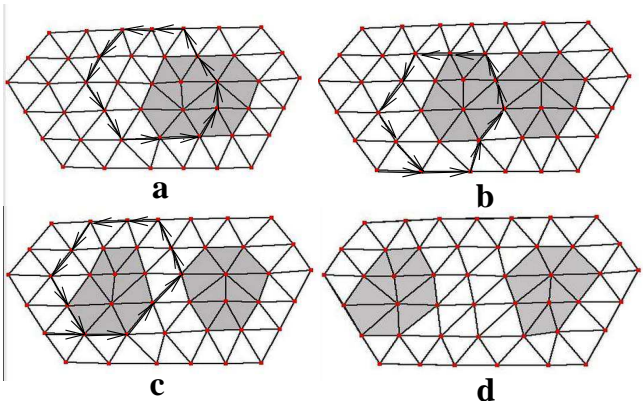}
\end{center}
\caption{The figures a -- d which we have drawn using the applet "voroglide"\cite{voro} show  four steps of separation of a type I dislocation pair, from a separation of $a_0$ to  $4a_0$. The shaded regions show the $5-7$ disclination pairs constituting the dislocations. Burger's circuits are shown in a -- c. Note that for separations $\geq 2 a_0$ separate Burger's circuits around  each disclination pair give rise to non-zero Burger's vectors, giving the dislocations their individual identity. This shows that the minimum meaningful separation 
between dislocation cores $r_{min} = 2 a_0$.}
\label{mv-dislo}
\end{figure}

\begin{figure}[t]
\begin{center}
\includegraphics[width=7.0cm]{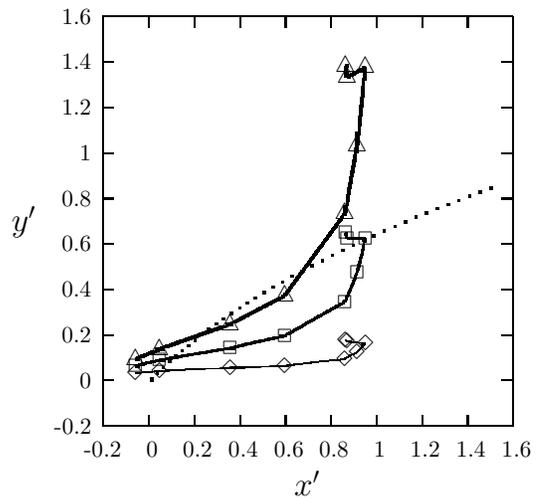}
\end{center}
\caption{Similar to Fig.\ref{flow}. The initial conditions $x'_0$ and
$y'_0$ are plotted as a function of $\beta V_0$. The different data sets
are created for different values of $r_{min}$. The symbols mean the
following :  $\Diamond$ denotes data for $r_{min} = a_0$, $\Box$ denotes that 
for $r_{min} = 2 a_0$ and $\triangle$ denote data for $r_{min} = 3 a_0$.
The dotted line denotes the non- linear separatrix. 
}
\label{comp_rmin}
\end{figure}
\vskip .2cm
\noindent
One aspect of our study which stands out is the exceptionally better
agreement of our results with previous simulations for hard disks as opposed
to systems with soft potentials like the soft disks and the DLVO. 
This could, in principle, be due either (a) to the failure of the RG 
equations used by us or some other assumptions in our 
calculations (b) or to unaccounted finite size effects in earlier simulations.
While it is difficult to estimate the effect of (a) since RG equations to 
higher orders in $y$ are unknown at present, we may be able to motivate an 
estimation for (b). In order to explain the  
discrepancy in the positions of the phase boundaries, we need to go into 
some details of how the phase diagrams were
obtained in the earlier simulations. In these simulations 
\cite{lif-hd,lif-sd1,lif-sd2,lif-dlvo} the phase boundaries were obtained 
from the crossing of the order parameter cumulants \cite{bincu,lan-bin} for 
various coarse graining sizes.
The system sizes simulated in these studies are the same ($N = 1024$). However,
the range of interaction differs. 
To obtain an objective measure we define the range of the potentials $\xi$  
as that at which the interaction potential $\phi$ is only 
$1 \%$ of its value at the lattice parameter. 
In units of lattice parameter, we obtain, for soft disks  $\xi = 1.47$ 
and for the DLVO potential $\xi= 1.29$ at typical screening
of $\kappa a_s = 15$. By definition, for hard disks $\xi=1$.
The particles within the range of the potential 
are highly correlated and we calculate the number $N_{corr}$ of such 
independent bare {\em uncorrelated particles} within the full system size. 
$N_{corr}$ takes the values $N_{corr} =  1024,~473.88,~615.35$ for hard disks, 
soft disks and the DLVO potential respectively.
Since the effective system sizes are smaller for the soft potentials, finite 
size effects are expected to be larger. In this connection, it is of interest 
to note that in the same publications \cite{lif-hd,lif-sd1,lif-sd2,lif-dlvo}
a systematic finite size 
analysis showed that the phase diagrams shift towards higher (lower) 
density (kappa) for hard and soft disks (DLVO).
A look at Fig.\ref{soft-phdia} and \ref{dlvo-phdia} should convince the 
reader that such a shift would actually make the agreement with our results 
better. We emphasize here that our present restricted simulations are virtually 
free of finite size effects since the system does not undergo 
any phase transition.  

\vskip .2cm
\noindent
Before we end this section,
we discuss the reasons behind the particular choice of $r_{min}$ that we
made throughout this manuscript. 
In practice, it is possible to  give individual identification to a dislocation only when Burger's vector separation within a pair is 
$\geq 2 a_0$ (Fig.\ref{mv-dislo}) {\em i.e.} $r_{min} = 2a_0$. 
For $r\geq 2 a_0$ Burger's loops can be drawn around each $5-7$ 
disclination pair (Fig.\ref{mv-dislo}) giving rise to a 
non-zero Burger's vector. In unrestricted simulations and in experimental 
situations after a disclination quartet is formed, they get separated out 
and the easy direction of separation is the glide 
direction which is parallel to the Burger's vector. In Fig.\ref{mv-dislo} we 
show four steps of separation of such a dislocation pair of type I.
After motivating $r_{min} = 2a_0$ we show, in Fig.\ref{comp_rmin}, the 
three sets of initial values corresponding to $r_{min} = a_0,~2a_0,~3a_0 $
along with the non-linear separatrix at $\eta = .7029$ of hard disk system.
$r_{min} = a_0$ predicts the system to be in solid phase for any arbitrarily 
small amount of external potential and to melt at larger $\beta V_0$. This 
behaviour contradicts physical expectation that the melting density at 
$\beta V_0 = 0$ has to be larger than that at  $\beta V_0 = \infty$. 
On the other hand, while $r_{min} = 3a_0$ does not produce any unphysical 
prediction, it shrinks the region of re-entrance in the $\beta V_0$ direction. 
Therefore $r_{min} = 2a_0$ is the minimum value for $r_{min}$ that could be 
chosen to produce physically meaningful results and this choice remains in 
closest agreement with simulation data. 
\noindent
\section{Conclusion}
\label{conclusion}
We have presented a complete numerical renormalization group scheme to
calculate phase diagrams for 2-d systems under a commensurate modulating 
potential. We have used FNR theory along with this scheme to calculate
phase diagrams for three different systems, namely, the hard disks, the 
DLVO and the soft disks. In all the cases we have found laser induced freezing
followed by a re-entrant laser induced melting. 
We show that the re-entrance behavior is built into the `bare' quantities 
themselves. We find extremely good agreement with earlier simulation results. 
In particular the phase diagram for hard disk comes out to be exactly the same 
as found from one set of earlier simulations\cite{lif-hd}.To obtain the correct 
phase diagram, however, flow equations
need to be correct at least upto next to leading order terms in the dislocation
fugacity. Our results, especially for small potential strengths, is 
particularly sensitive to these terms. Cross-over effects
from zero potential KTHNY melting transition are also substantial at small
values of the potential.

\vskip 0.2cm
\noindent
\acknowledgments
The authors thank Peter Nielaba, Wolfram Strepp, Abhishek Chaudhuri, 
Erwin Frey, Abhishek Dhar, Madan Rao and Yacov Kantor for useful discussions; 
D. C. thanks C.S.I.R., India, for a fellowship. 
Financial support by DST grant SP/S2/M-20/2001  is gratefully acknowledged.


\vskip 1.5cm

\begin{thebibliography}{10}

\bibitem{kt}
J.~M. Kosterlitz and Thouless, J. Phys. C {\bf 6},  1181  (1973).

\bibitem{surajitxy}
S. Sengupta, P. Nielaba, and K. Binder, Europhys. Lett. {\bf 50},  668  (2000).

\bibitem{sura-hdmelt}
S. Sengupta, P. Nielaba, and K. Binder, Phys. Rev. E {\bf 61},  6294  (2000).

\bibitem{joseph}
M.~S. Rzchowski, S.~P. Benz, M. Tinkham, and C.~J. Lobb, Phys. Rev. B {\bf 42},
   2041  (1990).

\bibitem{stripe}
S.~N. Coppersmith {\it et~al.}, Phys. Rev. Lett. {\bf 46},  549  (1981).

\bibitem{chowdhury}
A. Chowdhury, B.~J. Ackerson, and N.~A. Clark, Phys. Rev. Lett. {\bf 55},  833
  (1985).

\bibitem{wei}
Q.-H. Wei, C. Bechinger, D. Rudhardt, and P. Leiderer, Phys. Rev. Lett. {\bf
  81},  2606  (1998).

\bibitem{xy-expt-1}
F. Huang, M.~T. Kief, G.~J. Mankey, and R.~F. Willis, Phys. Rev. B {\bf 49},
  3962–3971  (1994).

\bibitem{xy-expt-2}
W. Durr {\it et~al.}, Phys. Rev. Lett. {\bf 62},  206–209  (1989).

\bibitem{fight}
{\em Ordering in Two Dimensions}, edited by S.~K. Sinha (North-Holland,
  Amsterdam, 1980).

\bibitem{kthny1}
D.~R. Nelson and B.~I. Halperin, Phys. Rev. B {\bf 19},  2457  (1979).

\bibitem{kthny2}
A.~P. Young, Phys. Rev. B {\bf 19},  1855  (1979).

\bibitem{strand}
K.~J. Strandburg, Phys. Rev. B {\bf 34},  3536  (1986).

\bibitem{nelson-dg}
D.~R. Nelson,  in {\em Phase Transitions and Critical Phenomena}, edited by C.
  Domb and J. Lebowitz (Academic Press, New York, 1983), Vol.~7, p.\ 1.

\bibitem{clock}
M.~S.~S. Challa and D.~P. Landau, Phys. Rev. B {\bf 33},  437  (1986).

\bibitem{dyn-2dmelt}
A. Zippelius, B.~I. Halperin, and D.~R. Nelson, Phys. Rev. B {\bf 22},  2514
  (1980).

\bibitem{mylif}
D. Chaudhuri and S. Sengupta, Europhys. Lett. {\bf 67},  814  (2004).

\bibitem{myerrlif}
D. Chaudhuri and S. Sengupta, Europhys. Lett. {\bf 68},  160  (2004).

\bibitem{jcdlvo}
J. Chakraborti, H.~R. Krishnamurthy, A.~K. Sood, and S. Sengupta, Phys. Rev.
  Lett. {\bf 75},  2232  (1995).

\bibitem{cdas1}
C. Das and H.~R. Krishnamurthy, Phys. Rev. B {\bf 58},  R5889  (1998).

\bibitem{cdas2}
C. Das, A.~K. Sood, and H.~R. Krishnamurthy, Physica {\bf A 270},  237  (1999).

\bibitem{cdas3}
C. Das, P. Chaudhuri, A.~K. Sood, and H.~R. Krishnamurthy, Current Science {\bf
  80},  959  (2001).

\bibitem{lif-hd}
W. Strepp, S. Sengupta, and P. Nielaba, Phys. Rev. E {\bf 63},  046106  (2001).

\bibitem{lif-dlvo}
W. Strepp, S. Sengupta, and P. Nielaba, Phys. Rev. E {\bf 66},  056109  (2002).

\bibitem{lif-sd1}
W. Strepp, S. Sengupta, M. Lohrer, and P. Nielaba, Computer Physics
  Communications {\bf 147},  370  (2002).

\bibitem{lif-sd2}
W. Strepp, S. Sengupta, M. Lohrer, and P. Nielaba, Mathematics and Computers in
  Simulation {\bf 62},  519  (2003).

\bibitem{jay}
J. Chakraborti, H.~R. Krishnamurthy, and A.~K. Sood, Phys. Rev. Lett. {\bf 73},
   2923  (1994).

\bibitem{frey}
E. Frey, D.~R. Nelson, and L. Radzihovsky, Phys. Rev. Lett. {\bf 83},  2977
  (1999).

\bibitem{colbook}
I.~W. Hamley, {\em Introduction to {S}oft {M}atter: polymer, colloids,
  amphiphiles and liquid crystals} (Wiley, Cluchester, 2000).

\bibitem{bincu}
K. Binder, Phys. Rev. Lett. {\bf 47},  693  (1981).

\bibitem{lan-bin}
D.~P. Landau and K. Binder, {\em A Guide to Monte Carlo Simulations in
  Statistical Physics} (Cambridge University Press,, Cambridge, UK, 2000).

\bibitem{jaster}
A. Jaster, Physica A {\bf 277},  106  (2000).

\bibitem{morf}
D.~S. Fisher, B.~I. Halperin, and R. Morf, Phys. Rev. B {\bf 20},  4692
  (1979).

\bibitem{amit}
D.~J. Amit, Y.~Y. Goldschmidt, and G. Grinstein, J.~Phys. A: Math. Gen. {\bf
  13},  585  (1980).

\bibitem{kram-like}
This assumption is similar in spirit to assuming that a particle which reaches
  the saddle point in the Kramers barrier crossing problem would automatically
  cross the barrier \cite{kramers}  .

\bibitem{lif-note-1}
Note that the calculation of the bare fugacity from the dislocation probability
  is, we believe, more accurate that the procedure used in \cite{mylif}  .

\bibitem{elast}
O. Farago and Y. Kantor, Phys. Rev. E {\bf 61},  2478  (2000).

\bibitem{pot-impact}
In the presence of an external 1D modulating potential periodic in the $y$-
  direction the stress has contribution from another virial- like additive
  term, $-\frac{\beta{\rm d}^2}{S}\left<\sum_\lambda y^\lambda
  f_y^\lambda\right>$, where $y^\lambda$ is the $y$-component of position
  vector of particle $\lambda$. This contribution comes from the part of the
  free energy that involves higher energy (massive) excitations. For the
  elastic free energy which is lowest order in the displacement gradient
  (Eq.\ref{hamiltonian}) this part does not contribute towards the elastic
  constants, as the $x$- and $y$- component of gradient remain uncoupled. This
  extra term in stress remains a constant background without disturbing the
  elastic constants connected to the Young and shear modulus that corresponds
  to distortions of the system in the low energy directions. We therefore
  neglect this background in calculating stresses where from we obtain the
  elastic moduli.  .

\bibitem{dlvo1}
E.~J.~W. Verwey and Overbeek, {\em Theory of Stability of Lyophobic Colloids}
  (Elsevier, Netherlands, 1948).

\bibitem{dlvo2}
Derjaguin, B. V., Landau, and L. D., Acta Phys. Chim., USSR {\bf 14},  633
  (1941).

\bibitem{al}
B. Alder and T. Wainwright, Phys. Rev. {\bf 127},  359  (1962).

\bibitem{zo}
J. Zollweg, G. Chester, and P. Leung, Phys. Rev. B {\bf 39},  9518  (1989).

\bibitem{web}
H. Weber and D. Marx, Europhys. Lett. {\bf 27},  593  (1994).

\bibitem{rhyzov}
V. Ryzhov and E. Tareyeva, Phys. Rev. B {\bf 51},  8789  (1995).

\bibitem{hansen-macdonald}
J.~P. Hansen and I.~R. MacDonald, {\em Theory of simple liquids} (Wiley,
  Cluchester, 1989).

\bibitem{dielec}
This interaction in colloids is due to polarization of the dielectric colloidal
  particles by the electric field of the laser. Though experiments of
  Refs.\cite{chowdhury,wei} use charged colloids, the interaction of hard
  sphere colloids with lasers is similar  .

\bibitem{frenkel}
D. Frenkel and B. Smith, {\em Understanding Molecular Simulation}, 2nd ed.
  (Academic Press, New York, 2002).

\bibitem{lif-note-2}
In analysing Fig.3 we must keep in mind that we can calculate from our
  simulations only the probability of formation of a disclination quartet.
  While we can, perhaps, safely assume that if type I dislocations are
  involved, they will seperate out with unit probability, the same can not be
  said of type II dislocations. This means that the probability of type II
  dislocations could be much lower than what Fig.3 suggests.  .

\bibitem{bch-fr}
C. Bechinger and E. Frey, J.~Phys.: Condens. Matter {\bf 13},  R321   (2001).

\bibitem{voro}
http://www.pi6.fernuni-hagen.de/GeomLab/VoroGlide/index.html.en  .

\bibitem{kramers}
H.~A. Kramers, Physica {\bf 7},  284  (1940).

\end{thebibliography}
\end{document}